\documentclass[sigplan,screen]{acmart}

\setcopyright{acmlicensed}
\acmPrice{15.00}
\acmDOI{10.1145/3519939.3523720}
\acmYear{2022}
\copyrightyear{2022}
\acmSubmissionID{pldi22main-p495-p}
\acmISBN{978-1-4503-9265-5/22/06}
\acmConference[PLDI '22]{Proceedings of the 43rd ACM SIGPLAN International Conference on Programming Language Design and Implementation}{June 13--17, 2022}{San Diego, CA, USA}
\acmBooktitle{Proceedings of the 43rd ACM SIGPLAN International Conference on Programming Language Design and Implementation (PLDI '22), June 13--17, 2022, San Diego, CA, USA}

\usepackage{amsmath}
\usepackage{amsthm}
\usepackage{listings}
\usepackage{comment}
\usepackage{amsfonts}
\usepackage{algorithmic}
\usepackage{graphicx}
\usepackage{textcomp}
\usepackage{hyperref}
\usepackage{xcolor}
\usepackage{xcolor}
\usepackage{graphicx}
\usepackage{subcaption}
\usepackage{multicol}
\usepackage{listings}
\usepackage{color}
\usepackage{multirow}
\usepackage{multirow}
\usepackage{rotating}
 \usepackage{microtype}
\usepackage{comment}
\usepackage{balance}
\usepackage{enumitem}
\usepackage[normalem]{ulem}
\usepackage{microtype}

\usepackage{url}

\begin{document}

\renewcommand*{\thefootnote}{\fnsymbol{footnote}}

\newcommand{\tool}[1]{\textsc{Gocc}}
\newcommand{\lib}[1]{\texttt{optiLib}}
\newcommand{\cs}[1]{\textbf{CS}}
\newcommand{\mutex}[1]{\texttt{Mutex}}
\newcommand{\lupair}[1]{LU-pair}
\newcommand{\fastpair}[1]{\texttt{FastLock()/FastUnlock()}}
\newcommand{\luset}[1]{\texttt{LU-Set}}
\newcommand{\milind}[1]{{\color{red} milind: #1}}
\newcommand{\adam}[1]{{\color{orange} Adam: #1}}
\newcommand{\chris}[1]{{\color{blue} chris: #1}}
\newcommand{\tim}[1]{{\color{cyan} Tim: #1}}
\newcommand{\change}[2]{\sout{#1} #2}
\newcommand{\timchange}[2]{\tim{\change{#1}{#2}}}
\newcommand{\optiLock}{\texttt{OptiLock}}
\newcommand{\optiRWLock}{\texttt{OptiRWLock}}
\newcommand{\gwt}{\texttt{GWT}}
\newcommand{\lp}{lock-point}
\newcommand{\up}{unlock-point}
\newcommand{\lup}{LU-points}
\newcommand{\speedup}{10$\times$}
\newcommand{\Eqnum}[1]{(\ref{eqn:#1})}
\newcommand{\Eqn}[1]{Equation~\Eqnum{#1}}
\newtheorem{Definition}{Definition}[section]
\newtheorem{Theorem}{Theorem}
\newtheorem{Remark}{Remark}
\newcommand{\QED}{\begin{flushright}$\Box$\end{flushright}}
\newtheorem{Observation}{Observation}
\newtheorem{Rule}{Rule}
\newtheorem{Fact}{Fact}
\newtheorem{Lemma}{Lemma}
\newtheorem{Corollary}{Corollary}
\newtheorem{Lcorol}{Corollary}
\newtheorem{Example}{Example}
\newenvironment{Proof}{\noindent{\bf Proof: }}{\qed}

\title{A Study of Real-World Data Races in Golang}



\author{Milind Chabbi}
\orcid{nnnn-nnnn-nnnn-nnnn}             
\affiliation{
  \department{Programming Systems Group}              
  \institution{Uber Technologies Inc.}            
  \city{Sunnyvale}
  \state{CA}
  \country{USA}                    
}
\email{milind@uber.com}          

\author{Murali Krishna Ramanathan}
\orcid{nnnn-nnnn-nnnn-nnnn}             
\affiliation{
  \department{Programming Systems Group}              
  \institution{Uber Technologies Inc.}            
  \city{New York City}
  \state{NY}
 \country{USA}                   
}
\email{murali@uber.com}         

\begin{abstract}

The concurrent programming literature is rich with tools and techniques for data race detection.
Less, however, has been known about real-world, industry-scale deployment, experience, and insights about data races.
Golang (Go for short) is a modern programming language that makes concurrency a first-class citizen. 
Go offers both message passing and shared memory for communicating among concurrent threads.
Go is gaining popularity in modern microservice-based systems.
Data races in Go stand in the face of its emerging popularity.

In this paper, using our industrial codebase as an example, we demonstrate that Go developers embrace concurrency and show how the abundance of concurrency alongside language idioms and nuances make Go programs  highly susceptible to data races. 
Google's Go distribution ships with a built-in dynamic data race detector based on ThreadSanitizer.
However, dynamic race detectors pose scalability and  flakiness challenges; we discuss various software engineering trade-offs to make this detector work effectively at scale.
We have deployed this detector in Uber's 46 million lines of Go codebase hosting 2100 distinct microservices, found over 2000 data races, and fixed over 1000 data races, spanning 790 distinct code patches submitted by 210 unique developers over a six-month period.
Based on a detailed investigation of these data race patterns in Go, we make seven high-level observations relating to the complex interplay between the Go language paradigm and data races. 
\end{abstract}


\maketitle

\section{Introduction}

Uber\footnote{\url{http://www.uber.com}} caters to hundreds of millions of daily customers who depend on its real-time transportation and delivery services.
Uber's back-end software is a microservice~\cite{microserviceWWW} architecture running on several million CPU cores in many data centers.
Different microservices talk to one another via remote-procedure calls (RPC). 
Individual services are written in various programming languages such as Go, Java, Python, and NodeJS.
Uber has extensively adopted Go as a primary programming language for developing these microservices.
Uber's Go monorepo~\cite{monorepo} comprises about 46 million lines of code and contains approximately 2100 unique Go services.

Go is one of the de facto languages  for implementing microservices. 
There are several benefits of Go~\cite{golangEasyGC, golangPerf, golangPerf2, golangPerf3, golangPerf4, golangPerf5}. 
Concurrency is a first-class citizen in Go.
Go requires minimal garbage collection tuning effort, unlike other languages such as Java.
Go is compiled to native binary, making it performant.
Go embraces a {\em minimalist} approach with Python-like syntax making it easy to learn and well-suited for rapid code development.
Finally, Go has rich tooling, library support, and a thriving open-source community.

Go makes it easy to express asynchronous tasks in programs.
Prefixing a function call with the \texttt{go} keyword runs the call asynchronously in the address space of the calling process.
These asynchronous function calls in Go are called \emph{goroutines}~\cite{golangGoroutine}.
Developers hide latency (e.g., IO or RPC calls to other services) by creating goroutines within Go programs.
Goroutines are considered ``lightweight'', and the Go runtime context switches them on the operating-system (OS) threads. 
Go programmers use goroutines liberally both for symmetric and asymmetric tasks.
Two or more goroutines can communicate data via message passing (channels~\cite{golangChan}) or shared memory.
Shared memory is the most commonly used means of data communication in Go~\cite{tu-asplos19}.

Go's memory model~\cite{sorin2011primer} is loosely defined and not standardized; it is claimed to be ``like DRF-SC''~\cite{gomemmodelCox}.
Go memory model defines a happens-before-based partial ordering among the events in different goroutines~\cite{gomemmodel}.
Synchronization events establish a happens-before partial order between two or more participating goroutines.
For example, a \emph{send} event on a channel by a goroutine is considered to happen before the corresponding \emph{receive} event on the same channel in another goroutine;
an  \emph{unlock} event on a mutex by a goroutine is considered to happen before the subsequent \emph{lock} event in another goroutine on the same mutex object.
A data race occurs in Go when two or more goroutines access the same datum, at least one of them is a write, and there is no ordering between them~\cite{gomemmodel, goracedetect, BoehmMiscompile}. 



Outages caused by data races in Go programs are a recurring and painful problem in our microservices.
These issues have brought down our critical customer-facing services for hours together, causing inconvenience to our customers and impacting our revenue.
Our developers have also found it hard to debug data races and sometimes resorted to fixing them by conservative strategies such as eliminating the concurrency altogether in suspicious code regions.

Data races form a popular category of bugs in shared-memory systems (including Go) and have been the focus of many proposals to detect them in the last three decades~\cite{BoehmCPP, JavaSpecWWW}. 
Two techniques for data race detection are popular --- static analysis~\cite{voung2007relay, racerd, EnglerRacerX, chord, errorprone, coverity} and dynamic analysis~\cite{SchonbergDatarace, JohnOntheflyRace, FengCilkRace, YiziOpenMp, RaghavanRace, DinningDatarace, fasttrack, djit, cp-racedetection,bond-oopsla15,hb-oopsla18, eraser,samak-pldi15}.
Google's Go distribution ships with a built-in dynamic data race detector~\cite{goraceDetectorWWW} based on ThreadSanitizer~\cite{SerebryanyThreadSanitizer}, which integrates lock-set~\cite{eraser} and happens-before~\cite{fasttrack, djit} algorithms to report races.
The cost of race detection varies by program, but for a typical program, memory usage increases by $5\times$-$10\times$ and execution time grows by $2\times$-$20\times$~\cite{goraceDetectorWWW}.
Additionally, the compilation time also increases by $\sim$$2\times$.

In this paper, we discuss deploying  Go's default dynamic race detector to continuously detect data races in Uber's Go development environment. 
We elaborate on our learning from this deployment by providing a detailed overview of the common patterns for introducing data races in Go programs and the facilities inherent in the language infrastructure that eases the introduction of races into production code. Our analysis is based on more than {\em one thousand data races} fixed by more than two hundred engineers.


While there have been many algorithms to perform dynamic race detection, there is a significant gap in the know-how on deploying dynamic analysis in a real-world setting. 
Given the intrinsic non-determinism associated with dynamic race detection, deploying it as part of  continuous integration~\cite{continuous-integration} is impractical. On the other hand, deploying it as a post-facto detection process introduces complexities in reporting races without duplication and challenges in determining the correct owner. We elaborate on the design choices for our deployment. 

We use more than 100K Go unit tests to exercise the code and detect data races in our repository. 
In about six months, our continuous monitoring system has detected over 2000 data races in the repository; during this time 210 unique developers fixed more than 1000 data races (corresponding to 790 unique patches).
The remaining ($\sim$$1000$) data races are being actively worked upon.
The system detects about five new data races every day in newly introduced code.

Our analysis of the detected and fixed races from this period reveals a surprising result that suggests that there are aspects unique to Go, beyond common concurrency bug type patterns, that introduce these races. We find that several Go language design nuances such as t
ransparent capture-by-reference of free variables~\cite{golangCptureblog},
named return variables~\cite{golangNamedResult},
deferred functions~\cite{golangDefer}, 
ability to mix shared memory with message passing~\cite{golangChan}, 
indistinguishable value vs. pointer semantics~\cite{golangPtrVal, golangCptureblog}, 
extensive use of thread-unsafe built-in map~\cite{effectiveGoMaps}, 
flexible group synchronization~\cite{waitgrp}, 
and confusing slice~\cite{effectiveGoSlices} semantics when combined with the simplicity of introducing concurrency via goroutines make Go highly susceptible to data races.


\subsection{Technical Contributions}
We make the following technical contributions in this paper. 
\begin{enumerate}
    \item Discuss the concurrency characteristics of Go programs in production to motivate the need for effective solutions to detect and eliminate data races in Go.  
    \item Identify technical gaps with deploying dynamic race detection in a production system to continuously detect races and scale to support thousands of developers.
    \item Perform an elaborate, first-of-a-kind study of various data race patterns found in our Go programs.
    This study is conducted by analyzing over 1000 data races fixed by 210 unique developers over a six-month period. 

\end{enumerate}

\section{Concurrency in Go Services}
In this section, we provide a few key static and dynamic concurrency-related characteristics of our Go programs and compare them with programs in other languages to highlight some of Go's distinguishing features. 


Most of our Go programs reside in a single repository (monorepo~\cite{monorepo}) of about 46 million lines of Go code (MLoC) hosting 2100 unique microservices. In contrast, our Java monorepo constitutes 19 MLoC and  hosts 857 microservices.
As a rough approximation of the use of concurrency, we counted the number of concurrency creation constructs and synchronization constructs in these repositories and made the following observation:

\begin{Observation}
Developers using Go employ significantly more concurrency and synchronization constructs than in Java.
\end{Observation}

Table~\ref{tab:syncConstructs} shows the number of concurrency creation, point-to-point synchronization, and group communication constructs found  in our Go and Java monorepos.
For concurrency creation, we looked for the \texttt{.start()} construct in Java and the \texttt{go} construct in Go\footnote{the exact regular expressions are more involved.}.
For point-to-point synchronization, we searched for constructs related to \texttt{lock}, \texttt{unlock}, \texttt{acquire}, \texttt{release}, and \texttt{synchronized} in Java and searched for \texttt{Lock}, \texttt{Unlock}, \texttt{RLock}, \texttt{RUnlock}, and the channel (\texttt{<-}) constructs in Go.
For group communication, we looked for the instances of \texttt{CyclicBarrier}, \texttt{CountDownLatch}, and \texttt{Phaser} in Java and looked for the instances of \texttt{WaitGroup}~\cite{waitgrp} in Go.
While this look-up is coarse-grained and imperfect, it sheds light on significant quantitative differences in the use of concurrency constructs in the two languages.
Java code has 219 thread creation constructs per MLoC, whereas the same for Go is 250 per MLoC, which is not significantly different.
However, the Java code has 203 point-to-point synchronization constructs per MLoC, whereas the same for Go is 754.2 per MLoC, which is $3.7\times$ higher.
Similarly, the Java code has 55.9 group synchronization constructs per MLoC, whereas the same for Go is 104.2 per MLoC, which is $1.9\times$ higher.

\begin{table}[]
    \caption{Use of concurrency and synchronization constructs in Java vs. Go monorepo.}
\footnotesize
    \centering
    \begin{tabular}{c|c|c|c}
       Feature & Subfeature & Java  & Go  \\ \hline
        LoC & - & 19 Million & 46 Million \\ \hline 
        services & - & 857 & 2100 \\ \hline  \hline
        concurrency creation & - & 4162 & 11515 \\ \cline{2-4} 
        & \textbf{total/MLoC} & \textbf{219.1} & \textbf{250.3} \\ \hline \hline
        point-to-point  & synchronized & 2378 & - \\ 
        communication &  acquire+release & 652 & - \\
        &  lock+unlock & 624\footnote{includes reader-writer locks} &  19062\\
        &  rlock+runlock & - & 5511 \\ 
        &  channel send/recv & - & 10120 \\ \cline{2-4}
        & \textbf{total/MLoC} & \textbf{203} & \textbf{754.2} \\ \hline \hline
         Group  &  Latch/Barrier/Phaser & 1007 & - \\ 
       communication & WaitGroup & - & 4795 \\ \cline{2-4}  
        & \textbf{total/MLoC} & \textbf{55.9} & \textbf{104.2}\\ \hline
    \end{tabular}
    \label{tab:syncConstructs}
\end{table}

\begin{Observation}
Developers using Go for programming microservices expose significantly more runtime concurrency than other languages such as Java, Python, and NodeJS used for the same purpose.
\end{Observation}

\begin{figure}
    \centering
    \includegraphics[width=\linewidth]{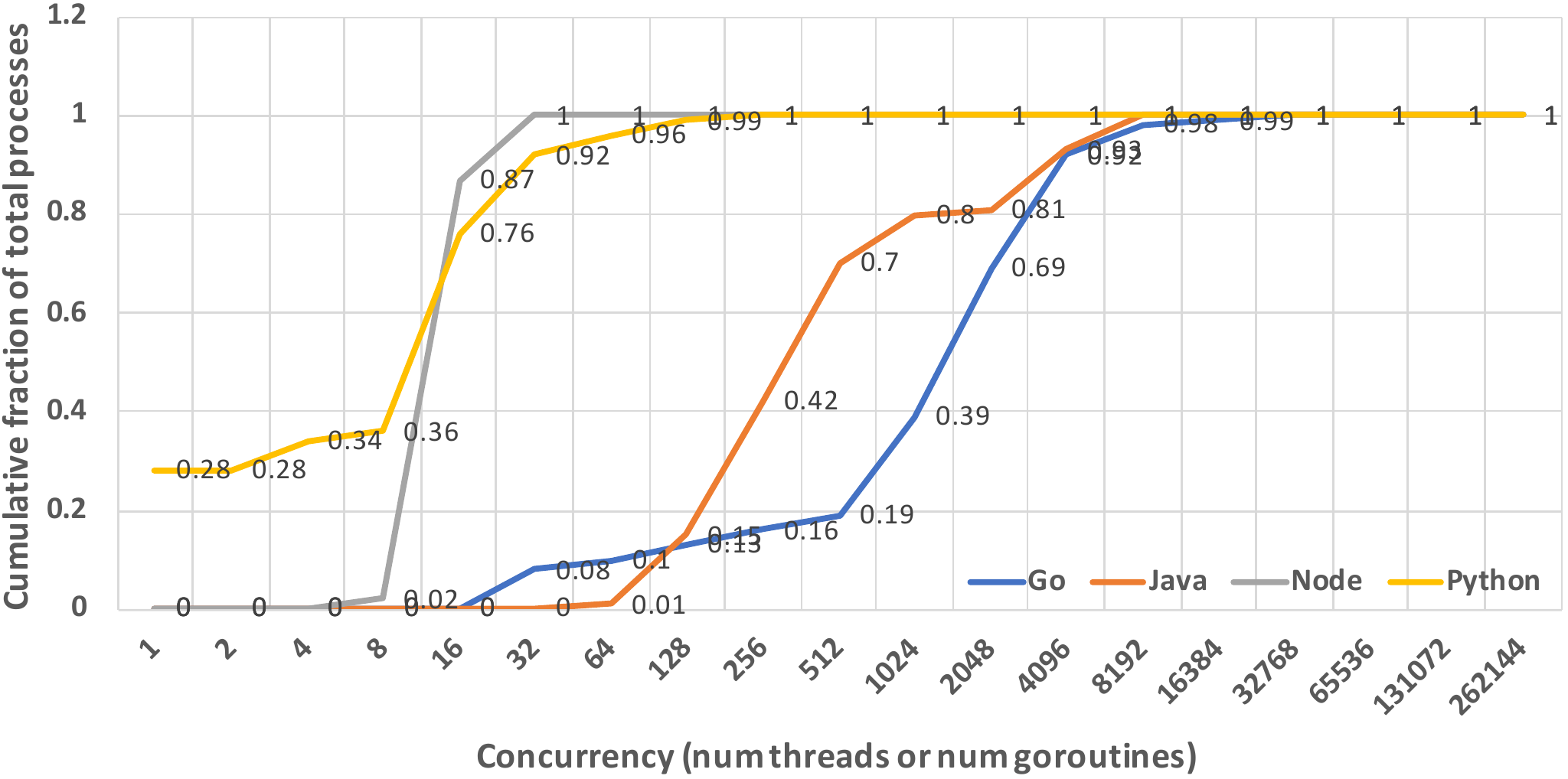}
    \caption{Cumulative frequency distribution of concurrency within programs of different languages.}
    \label{fig:concurrencyHisto}
\end{figure}
To substantiate this fact, we scanned our data centers and counted the number of threads in the service instances (processes) running on each machine.
For Go, we used the \texttt{pprof}~\cite{pprof} profiler to count the number of goroutines in each process. Our scan included 130K Go processes, 39.5K Java processes, 19K Python processes, and 7K NodeJS processes.
Go programs are most widely deployed, and the next one is Java.
Figure~\ref{fig:concurrencyHisto} shows the fraction of processes at different concurrency levels for each language, plotted as a cumulative frequency distribution graph.
The x-axis plots the concurrency level on a log scale.
The y-axis is the cumulative fraction of processes of each language.
The x value(s), where the line rises sharply, represents the common concurrency level for the specific language.
A line rising more on the higher value of x has a higher average concurrency.

From the figure, we note that NodeJS typically has 16 threads.
Python typically has less than 16-32 threads.
The second most used language in our fleet, Java, often has between 128-1024 threads; about 10\% of cases have 4096 threads, and 7\% have 8192 threads.
In contrast, typically, Go processes have 1024-4096 goroutines; about 6\% of processes contain 8102 goroutines. 
The max reaches at about 130K goroutines with contributions to the cumulative sum along the intermediate steps.
Put another way, the 50\% percentile of the number of threads is 16 in NodeJS, 16 in Python, 256 in Java, and 2048 in Go.
This data empirically demonstrates that Go programs, at runtime, expose a lot of concurrency --- about $8\times$ more than Java.

Since higher concurrency can introduce more concurrency bugs, it is only natural to expect more data races in Go, especially when the language does not have a built-in mechanism to avoid data races, unlike languages such as Rust~\cite{rustWWW, qin-pldi2020}.



\section{Deploying Dynamic Data Race Detector}
In this section, we provide a brief background on dynamic data race detection, discuss the design challenges of its deployment in an industrial setup, elaborate on our deployment design, and finally conclude with the impact of rolling this out to the engineering community internally. 

\subsection{Dynamic Data Race Detection}
A data race happens when a shared memory location is accessed {\em concurrently} by more than one thread and at least one of the accesses is a write access. The absence of a happens-before (HB) relation between the accesses results in the accesses being considered concurrent. 

Executing a program and analyzing its execution, by instrumenting the shared memory accesses and synchronization primitives, is used to detect underlying races. The execution of unit tests in Go that spawn multiple goroutines is a good starting point for dynamic race detection. Go has a built-in race detector that can be used to instrument the code at compile time and detect the races during their execution.  

\sloppy
Internally, the Go race detector uses the ThreadSanitizer~\cite{SerebryanyThreadSanitizer} runtime library. ThreadSanitizer uses a combination of lock-sets~\cite{eraser} and HB~\cite{djit, fasttrack} based algorithms to report races. 
The lock-sets algorithm tracks the set of locks held before every memory access. 
If the set intersection of lock-sets associated with the different accesses of a shared memory location across goroutines is empty, it might indicate a data race. 
The lock-sets algorithm does not track the HB relations and it may include races that may never manifest in practice. 

For higher precision, the Go race detector also uses vector clocks~\cite{lamportClocks} to track the HB relation between shared memory accesses.
Vector clocks are expensive both in space and time since they track the relation between the timestamps of accesses. We defer the reader to the vast literature that discusses these approaches elaborately. 
Although the scheduling of goroutines on underlying OS-threads is non-deterministic in Go, since the race detector maintains a vector clock per goroutine, the data race detector correctly reports a data race whether or not the goroutines ran on the same OS thread in a given execution.

There are three attributes one needs to be aware of with respect to dynamic race detection: 
\begin{enumerate}
    \item It may not report all races in the source code as it is dependent on the analyzed executions.
    \item The detected set of races depend on the thread interleavings and can vary across multiple runs, even though the input to the program remains unchanged.
    \item A reported race can be ``benign''~\cite{NarayanasamyBenignRace} and need not lead to a failing execution. 
\end{enumerate}

\subsection{Deployment Challenges}
We now discuss the challenges associated with deploying the dynamic race detector in a large codebase quickly and ensure that the appropriate developers fix these races. 

\subsubsection{When to deploy?} When code changes are sent for code review (e.g., a pull request (PR)~\cite{pull-requests}), it is common practice to run many low-cost static analysis checks~\cite{errorprone} along with executing relevant unit tests~\cite{predictive-test-selection}. This approach ensures that the problematic code changes are detected early and are not merged into the main branch of the source.  
This process has the following benefits:
\begin{enumerate}
\item Defects are fixed before they are pushed to production,
\item Defect ownership is straightforward as the author of the PR becomes responsible for fixing the bug, and
\item The time taken to fix the issue is comparatively lower as the author has sufficient context on the changes. 
\end{enumerate}

We explore the following three possibilities on when to deploy a dynamic data race detector. We ultimately adopt the last one --- generating bug reports based on data race detection in our workflow also referred to as ``post-facto data race bug reporting''.

\vspace{0.2cm}
\noindent \textbf{[Option I] Block the PR based on dynamic race detection:} An obvious question is whether a dynamic race detector can also be used as part of this workflow, referred to as Continuous Integration (CI)~\cite{continuous-integration}. In other words, when the PR is sent for review, should dynamic race detection also be performed to prevent potential data races from being introduced into the code? We observe that dynamic race detection is a misfit during this phase for the following reasons: 
\begin{enumerate}
    \item Dynamic race detection is non-deterministic~\cite{racefuzzer} as the paths followed by various executions may be different due to the thread interleavings. The flakiness may result in races being dormant in the PR where the bug is introduced and being detected in a later, unrelated PR and impacting a different author. 
    \item High overheads ($2\times$-$20\times$~\cite{goracedetect}) of dynamic data race detection impact the turnaround time for landing changes into the main branch. Reducing the overheads by instrumenting only the modified libraries will not be helpful as this affects overall analysis precision (e.g., impacts the correctness of HB relations that may happen in another library or the runtime).  
    \item The presence of pre-existing races complicates the deployment as it can overwhelm a developer who may not have been responsible for the data races.  
\end{enumerate}

A low overhead dynamic data race detector~\cite{literaceMarino, ProRaceZhang} is not a panacea for this situation; the non-determinism of detected races prevents us from adopting it at a PR time.

\vspace{0.2cm}
\noindent \textbf{[Option II] Add warnings to a PR based on dynamic race detection:} Given the challenges of blocking a PR based on the non-determinism of a dynamic data race detector, we considered making data races as non-blocking PR-time warnings. However, we did not pursue this direction for the following reasons:
\begin{enumerate}
\item Thousands of pre-existing data races make the warning report noisy,
\item Developers ignore warnings and become immune to it, and
\item The overhead of the dynamic race detector violates build-time SLAs.
\end{enumerate}

\vspace{0.2cm}
\noindent \textbf{[Option III] Generate bug reports based on dynamic race detection:}
Based on these considerations, our deployment strategy for dynamic race detection is to perform it periodically on a snapshot of code, post-facto, instead of incorporating it as part of the CI workflow.
Meaning, we do not prevent data races from being introduced into our codebase; we actively find and fix them.
This design assigns a detected data race as a defect to its owner using a heuristic. 

While incorrectly blocking a PR makes the build system flaky and hurts developer sentiments, accidentally assigning a data race defect to an incorrect owner is not a problem of the same magnitude. 
This is so because defects get triaged and eventually get reassigned to appropriate owners in a large software development environment.

Despite our current design choice, our experience with this deployment (and other non-CI-based analyses) shows that algorithms enabling the deployment of dynamic race detection during the CI workflow are more impactful, as they increase the likelihood of races being fixed.

\subsection{Post-facto Data Race Bug Reporting}

Deploying dynamic race detection periodically, post-facto, will broadly involve the following steps --- (a) clone the latest version of the repository, (b) perform dynamic race detection by executing all the unit tests in the repository, and (c) report all outstanding races by filing tasks to the appropriate bug owner.  A detected race report contains the following details:
\begin{enumerate}
    \item The conflicting memory address,
    \item Two call chains (aka calling contexts or stack traces) of the two conflicting accesses, and
    \item The memory access types (read or a write) associated with each access.
\end{enumerate}

This process exposes interesting design choices with building a practical developer workflow due to the challenges associated with the code constantly evolving, underlying non-determinism of dynamic race detection, and the frequent churn in the organization. 
We discuss these details below.

\subsubsection{Ensuring unique race reports} Each calling context comprises a list of function names and line numbers. 
Hashing all this information can help in unique identification of the race but can result in duplicated race reports, just because either a) the source line numbers change because of unrelated changes to the source file, or b) the order of two conflicting accesses is reversed.

To avoid such duplication and to ensure consistency in reporting across source code revisions, we devised a simple hashing technique that is relatively resilient to these common concerns.
We first ignore the source line numbers in both call chains, which takes care of unrelated code modifications within a function.
Second, we order the two call stacks lexicographically; meaning two call chains \texttt{P()->Q()->R()}  and \texttt{A()->B()->C()} are always ordered as \texttt{A()->B()->C() and P()->Q()->R()}, irrespective of the order in which the execution happened.

The flip side of ignoring the line numbers is that it suppresses  different data races sharing exactly the same two call chains but differing only in line numbers if at least one of them is already reported. This approach is not a significant limitation because we suppress a defect {\em iff} there is an active one with the same hash that is already open in our bug database. As soon as the open defect with the same hash is fixed, our system files another defect with the same hash (sharing the call chains), if it finds one.


On this topic, it must also be emphasized that the uniqueness of reporting is only about the manifested races and not the underlying root causes. For example, the same underlying root cause may result in different pairs of conflicting memory accesses (e.g., absence of a lock causing multiple shared data structures to race). 
See statistics in Section~\ref{sec:Experiences}.
Automatically triaging the root cause and reporting them uniquely is an interesting area of research worth exploring but is outside the scope of our current effort. 

\subsubsection{Determining assignee for a race} After detecting a race, our system needs to determine the developer who will be responsible for fixing the race. In the absence of the root cause, the potential list of assignees is restricted to the authors of the root and leaf nodes of the two call stacks. Even though the leaf nodes contain the race-inducing memory accesses, we choose to report it to the owner of the root nodes of the call stacks. We chose this option because the developers associated with the root nodes correspond to the author of code higher up in the call stack; these developers have a stake in the functional correctness of their code and are hence incentivized to eliminate a race and drive the issue to closure even if it is in a downstream library. 

Due to the churn in the organization and the presence of mass refactorings to source code, reporting a defect to the latest modifier of the code may result in an inaccurate assignment. Therefore, we consider a few additional statistics corresponding to a) the author(s) who frequently modify the associated source code, b) any metadata attached to the source describing the owning team, and c) the presence of the developer and their manager in the organization, etc. 
We found in our experience that attaching a log of how our algorithm arrived at the choice of the assignee and enumerating the potential set of assignees was useful to the developers, rather than simply assigning without explaining why the tool thinks the assignee should fix the data race.

Based on our experiences, we make the following remarks:

\begin{Remark}
      Design algorithms to enable dynamic race detection during Continuous Integration.  
\end{Remark}

\begin{Remark}
    Design algorithms to automatically root cause and identify  appropriate ownership for data races.  
\end{Remark}

\subsection{Deployment}

\begin{figure}[!t]
    \centering
    \includegraphics[trim={0 .3cm 0  0},clip,width=\linewidth]{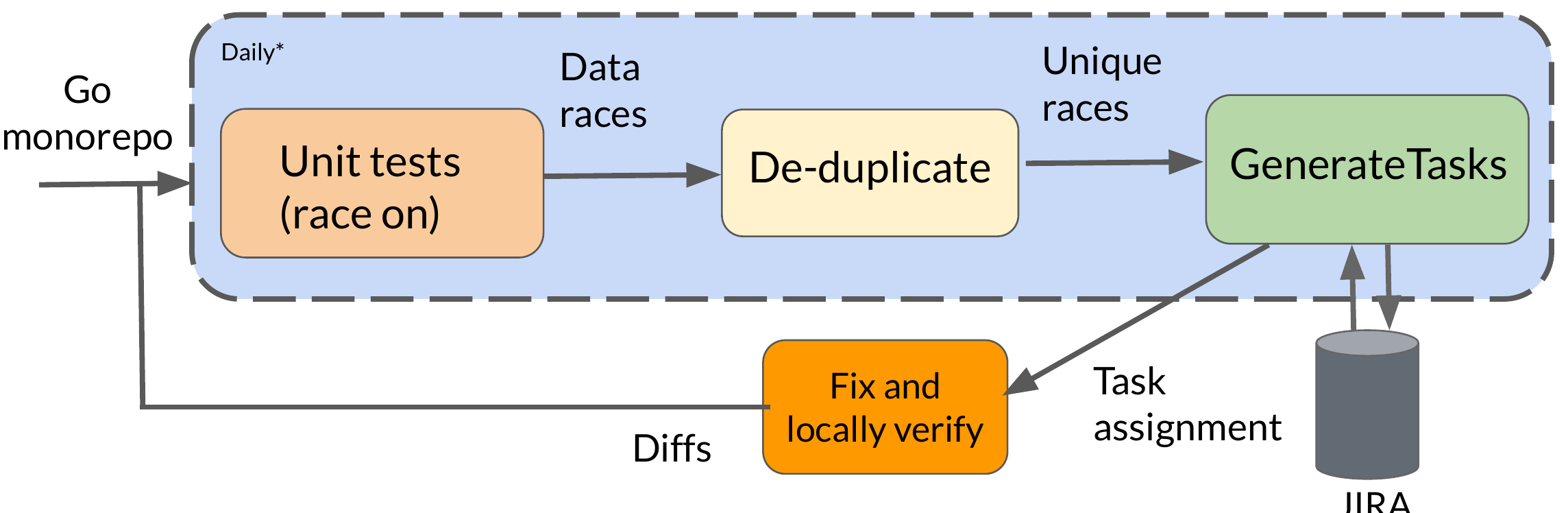}
    \caption{The architecture of data race detection.}
    \label{fig:arch}
\end{figure}

Figure~\ref{fig:arch} presents the architecture diagram of our deployment. Periodically (daily), the workflow executes all the unit tests with race enabled; this step involves compiling the entire program with instrumentation followed by executing tests.
The built-in dynamic race detector records encountered data races in each test to the standard error, which we capture for further processing.
The detected races are de-duplicated, and tasks are generated for the resulting unique races and filed to the appropriate developer(s). The task contains the details on the source version on which the race was detected, stack traces of two conflict accesses, and the necessary instructions to help the developer reproduce the underlying race. The fixes to the races follow standard development procedures, and the resultant changes are merged into the main branch of the source. 

\subsection{Experiences}
\label{sec:Experiences}
We rolled out this deployment in April 2021 and have collected data for about six months. Our approach has helped detect $\sim$2000 data races in our monorepo with hundreds of daily commits by hundreds of Go developers. 
Out of the reported races, 1011 races are fixed by 210 different engineers. In addition, we observe that there were 790 unique patches to fix these races, suggesting $\sim$78\% unique root causes. 

\begin{figure}[!t]
    \centering
    \includegraphics[width=\linewidth, height=3cm]{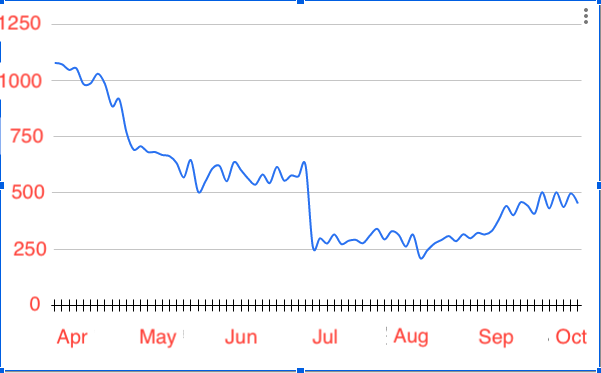}
    \caption{Total outstanding detected races vs. time.}
    \label{fig:race-time}
\end{figure}

We also collected the statistics for the total outstanding races over the last six months and report this data in Figure~\ref{fig:race-time}. In the initial phase (2-3 months) of the rollout,  we shepherded the assignees to fix the data races. The drop in the outstanding races is noticeable during this phase. Subsequently, as the shepherding was minimized, we notice a gradual increase in the total outstanding races. The figure also shows the fluctuations in the outstanding count, which is due to fixes to races, the introduction of new races, enabling and disabling of tests by developers, and the underlying non-determinism of dynamic race detection. After reporting all the pre-existing races, we also observe that the workflow creates about five new race reports, on average, every day. 
\begin{figure}[!t]
    \centering
    \includegraphics[width=.83\linewidth]{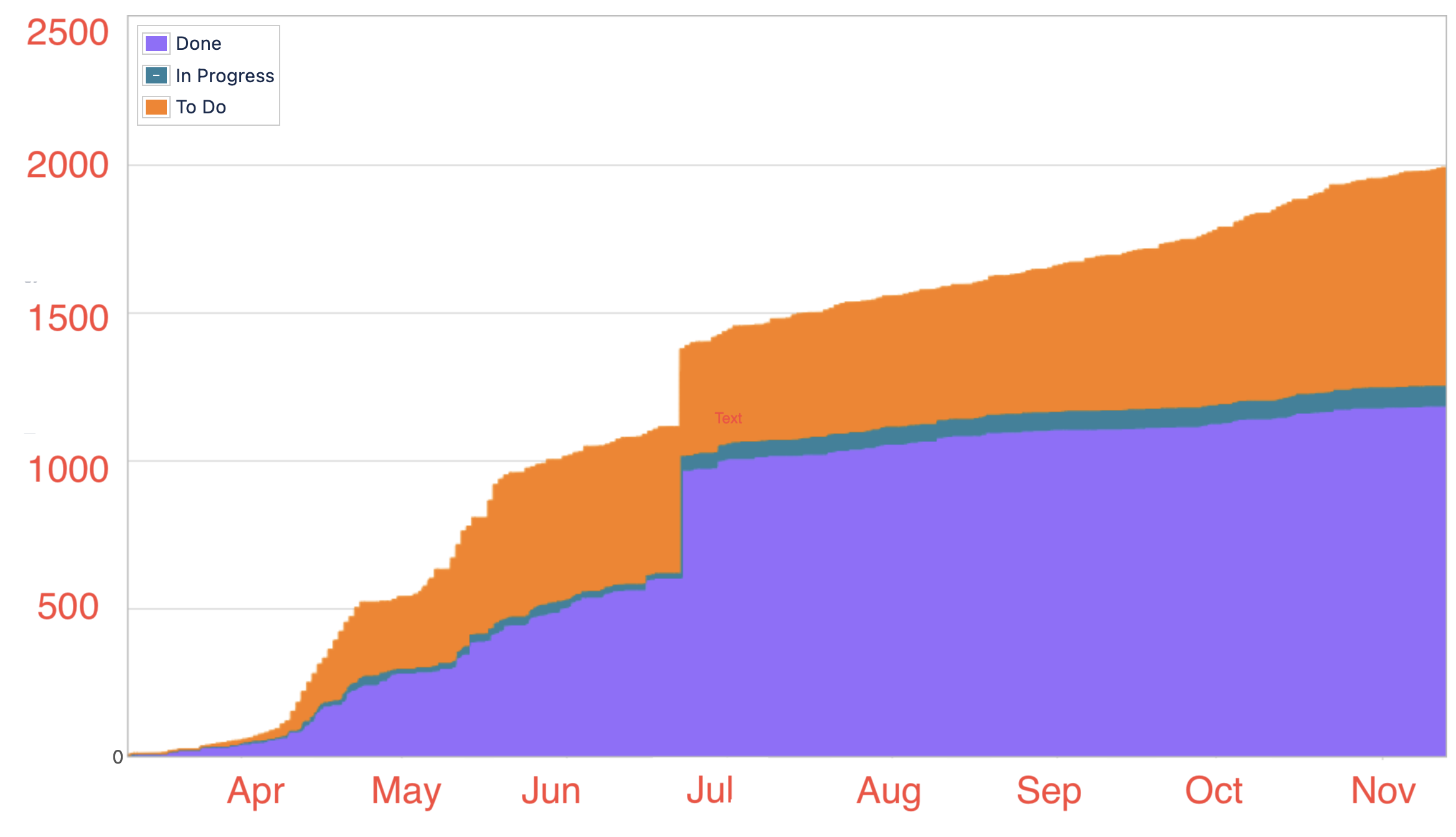}
    \caption{A timeline of data race issues found vs. fixed.}
    \label{fig:jira}
\end{figure}

Figure~\ref{fig:jira} also shows the activity on the tasks created by our deployment. 
There is a slow rise in reported races from April to June because we did not release all existing data races in a single day.
Instead, we slowly ramped up the number of data races we reported as we were fine-tuning our workflow to best suit our developers.
The sudden surge in July is a result of finally opening the flood gates, which also coincided with fixing a lot of bugs due to the authors a) driving the developers to fix the data races and b) themselves fixing several critical data races.
The initial phase resulted in a significant resolution of the created tasks. Subsequently, we observe that the gradient for the task creation is higher than that of task resolution because the authors disengaged from shepherding. 

We believe that the presence of race detection as part of a CI workflow will help address this problem by preventing new races from being introduced, apart from reducing the outstanding race count to zero.

In terms of the overhead of running our offline data race detector, we noticed that the 95th percentile of the running time of all tests without data race detection is 25 minutes, whereas it increases by $4\times$ to about 100 minutes with data race enabled.
Furthermore, in a survey taken by tens of engineers, about six months after rolling out the system, 52\% of developers found the system useful, 40\% of developers were not involved with the system, and 8\% of developers did not find it useful.

\section{Observations of Data Races in Go}
\label{sec:observations}
In this section, we detail several common patterns of data races we found in our investigation of over 1000 data races that were fixed. We have chosen, as our examples, the ones that either happen frequently or are unique and subtle to root cause. 
In the following examples, we have changed the function and variable names, reduced the code, and retained only the basic structure needed to understand the root cause of races.

\subsection{Essential Background on Go Syntax}
As a precursor to the rest of this section, we provide some essential background to the Go syntax~\cite{godevlist}.
Go is garbage collected, type-safe, and statically compiled.

In Go, the variable name precedes its type in a declaration, e.g., \colorbox{blue!30}{\texttt{var a int}}.
The syntax, \colorbox{blue!30}{\texttt{x := 10}}, both declares \texttt{x} as an integer via type inference and defines its value to be $10$.
Functions can return multiple values, similar to Python.
Arrays, declared as \colorbox{blue!30}{\texttt{var myArr [SIZE]int}}, are value types in Go and are passed by value, unlike C, C++, or Java.
Slices, declared \colorbox{blue!30}{\texttt{var mySlice []int}}, are dynamically growing/shrinking indexable data structure.
A slice in Go is a reference type, meaning it is heap-allocated.
Hash table is a built-in type. 
\colorbox{blue!30}{\texttt{var myMap map[int]string}} is a hash table with integer key and string value. Map is also a reference type in Go. 

Structs, like C++, are aggregate types in Go.
Go does not have classes. 
However, one can define methods on types.
A method is a function with a special ``receiver'' argument --- the object it is invoked on.
\colorbox{blue!30}{\texttt{func (m *myStruct) Foo()}} declares \texttt{Foo} as a method on \texttt{myStruct} type. 
Here, \texttt{m} is called the ``receiver''.
Notice that the receiver is a pointer (\texttt{*}) to \texttt{myStruct} in this example.

While C++, Java, and Python pass \colorbox{blue!30}{\texttt{this/self}} as a pointer to the method, in Go, it can be passed either as a pointer or a value depending on the type of the receiver.
One can  write a method on a value.
For example, \colorbox{blue!30}{\texttt{func (m myStruct) Foo()}} declares \texttt{Foo} as a method on \texttt{myStruct} value type, and the only difference is that the receiver here is a value type.
The method invocation in either case uses the same \colorbox{blue!30}{\texttt{obj.Foo()}} syntax.

Go has pointers, and they can be accessed by the standard address-of (\colorbox{blue!30}{\texttt{\&}}) operator on a value.
Go uses the same dot (\colorbox{blue!30}{\texttt{.}}) operator to access fields of a struct or invoke a method, whether on a value type (e.g., \texttt{myValue.Foo()}) or a pointer type (e.g., \texttt{myPtr.Foo()}).
The compiler infers whether a variable is a pointer or value and transparently arranges for dereferencing it or taking its address, as appropriate.

Go allows nesting functions (aka closures).
Nested functions are anonymous and capture  all the free variables \emph{by-reference.}
One can launch any function asynchronously by prefixing the call with the \colorbox{blue!30}{\texttt{go}} keyword.
Any function call with the prefix \colorbox{blue!30}{\texttt{defer}} delays its execution till the function's return.

We now discuss the various data race patterns by making an observation at the beginning of each subsection and explain the observation elaborately with an example.  All the example data races discussed in this section are available as an open-source artifact  at \url{https://zenodo.org/record/6330164}.

\subsection{Races due to Transparent Capture-by-Reference}
\begin{Observation}
Transparent capture-by-reference of free variables in goroutines is a recipe for data races.
\end{Observation}

Nested functions, aka closures, in Go \emph{transparently} capture all free variables~\cite{wikiFreeVar}  \emph{by-reference}.
The programmer does \emph{not} explicitly specify which free variables are captured in the closure syntax.
This mode of usage is different from, say, C++, where the programmer must explicitly specify every captured free variable as well as specify whether it is captured by-value or by-reference~\cite{cppLamdbaWWW}. 
In Java, all free variables are captured by-value~\cite{javaLamdbaWWW}.

Developers are often unaware that a variable used inside a closure is a free variable and captured by-reference, especially when the closure is large.
More often than not, Go developers use closures as goroutines.
As a result of capture-by-reference and goroutine concurrency, Go programs end up potentially having unordered access to free variables, unless explicit synchronization is performed.
The capture in a closure manifests itself in different forms, which we show in the following subsections.

\subsubsection{Loop index variable capture}
Listing~\ref{lst:rangeVarCapture} shows an example of iterating over a Go slice \texttt{jobs} and processing each element \texttt{job} via the \texttt{ProcessJob} function.
Here the developer wraps the expensive \texttt{ProcessJob} in an anonymous goroutine that is launched one per item.
However, the loop index variable \texttt{job} is captured by-reference inside the goroutine.
When the goroutine launched for the first loop iteration is accessing the \texttt{job} variable on line ~\ref{lst:indexcapture}, the  for loop will be advancing through the slice and updating the same  loop-index variable \texttt{job} (on line~\ref{lst:loopindex}) to point to the second element in the slice.
This situation causes a data race, which we have highlighted by the two black right triangles in the code.
This type of data race happens for value and reference types; slices, array, and maps; and read-only and write accesses in the loop body.
Go recommends a coding idiom to hide and privatize~\cite{goCommonMistakeWWW} the loop index variable in the loop body, which developers do not always follow.

\begin{figure}[t!]
\begin{lstlisting}[language=Go,label={lst:rangeVarCapture},caption=Loop index variable capture.]
@$\blacktriangleright$\label{lst:loopindex}@for _, job := range jobs {
  go func() {
@$\blacktriangleright$\label{lst:indexcapture}@  ProcessJob(job)
  }()
} // end for
\end{lstlisting}
\end{figure}

\subsubsection{Idiomatic \texttt{err} variable capture}
\begin{figure}[t!]
\begin{lstlisting}[language=Go,label={lst:errVarCapture},caption=Error variable capture.]
@\label{err:caller}@  x, err := Foo()
@\label{err:checkStart}@  if err != nil {
    ...
@\label{err:checkEnd}@  }
  
@\label{err:asyncStart}@  go func() {
@\label{err:defy}@    var y int
@$\blacktriangleright$\label{err:another}@   y, err = Bar()
    if err != nil {
      ...
    }
@\label{err:asyncEnd}@  }()

  var z int
@$\blacktriangleright$\label{err:yetanother}@ z, err = Baz()
  if err != nil {
    ...
  }
\end{lstlisting}
\end{figure}
Go allows and advocates multiple return values~\cite{golangMultiReturn} from functions.
It is common to return the actual return value(s) and an error object to indicate if there was an error. The actual return value is considered meaningful if and only if the error value is \texttt{nil}. 
In Listing~\ref{lst:errVarCapture}, Line~\ref{err:caller} shows an example invocation of a function \texttt{Foo}, which returns an integer and an \texttt{error} object.
It is a common practice to assign the returned \texttt{error} object to a variable named \texttt{err}, followed by checking for its \texttt{nil}ness as shown in lines~\ref{err:checkStart}-\ref{err:checkEnd}.
Since multiple error-returning functions can be called inside a function body, there will be several assignments to the \texttt{err} variable followed by the \texttt{nil}ness check each time, as shown on lines~\ref{err:another} and \ref{err:yetanother}.
Developers do not create a new error variable each time.

When developers mix this idiom with a goroutine, as shown in lines~\ref{err:asyncStart}-\ref{err:asyncEnd}, the \texttt{err} variable gets captured by-reference in the closure.
As a result, the accesses (both read and write) to \texttt{err} in the goroutine run concurrently with subsequent reads and writes to the same \texttt{err} variable in the enclosing function (or multiple instances of the goroutine), which causes a data race. In the example in Listing~\ref{lst:errVarCapture}, one write to \texttt{err} on line~\ref{err:another} and another write to the same \texttt{err} on line~\ref{err:yetanother} cause a data race.


\subsubsection{Named return variable capture}
Go introduces a syntactic sugar called ``named'' return values~\cite{golangNamedResult}.
The named return variables are treated as variables defined at the top of the function, whose scope outlives the body of the function.
A return statement without arguments, known as a "naked" return, returns the named return values. 
Go recommends using naked returns only in short functions because they can harm readability in longer functions.
However, developers use named returns in larger functions also.
In the presence of a closure, mixing normal (non-naked) returns with named returns or using a \texttt{defer}red~\cite{golangDefer} return in a function with a named return  is a risky proposition as we show below.

\begin{figure}[t!]
\begin{lstlisting}[language=Go,label={lst:namedVarCapture},caption=Named return variable capture.]
@\label{named:funcStart}@func NamedReturnCallee() (result int){
@\label{named:assign}@  result = 10
  if ... {
@\label{named:naked}@    return // this has the effect of "return 10" 
  }
@\label{named:goStart}@  go func(){
@$\blacktriangleright$\label{named:goRead}@   ... = result // read result
  }()
@$\blacktriangleright$\label{named:racyWrite}@ return 20 // this is equivalent to result=20
}

func Caller() {
@\label{named:caller}@  retVal := NamedReturnCallee()
}
\end{lstlisting}
\end{figure}

\paragraph{Normal return in a function with a named return:}
Listing~\ref{lst:namedVarCapture} shows mixing a normal return in a function
with a named return variable in the presence of a goroutine.
The function \texttt{NamedReturnCallee} returns an integer, and the return variable is named as \texttt{result} on line~\ref{named:funcStart}.
The rest of the function body can read and write to \texttt{result} without declaring it because of this syntax.
If the function returns at line~\ref{named:naked}, which is a naked return, due to the assignment \texttt{result=10} on line~\ref{named:assign}, the caller at line~\ref{named:caller} would see the return value of $10$.
The compiler arranges for copying \texttt{result} to \texttt{retVal}.

A named return function can also use the standard return syntax, as shown on line~\ref{named:racyWrite}.
This syntax makes the compiler copy the return value, $20$ in the return statement, to be assigned to the named return variable \texttt{result}.

Line~\ref{named:goStart} creates a goroutine, which captures the named return variable \texttt{result}.
In setting up this goroutine, even a concurrency expert might believe that the read from \texttt{result} on line~\ref{named:goRead} is safe because there is no other write to the same variable; the statement \texttt{return 20} on line~\ref{named:racyWrite} is, after all, a constant return and does not seem to touch the named return variable \texttt{result}.
The code generation, however, turns the \texttt{return 20} statement into a write to \texttt{result}, as previously mentioned.
Now suddenly, we have a concurrent read and a write to the shared \texttt{result} variable, a case of a data race.

\paragraph{Deferred functions in a named return:}
Yet another flavor of data races  happens when a named return is present alongside Go's \texttt{defer} construct.
\begin{figure}[t!]
\begin{lstlisting}[language=Go,label={lst:namedVarCaptureDefer},caption=Named return variable capture with a defer return.]
func Redeem(request Entity) (resp Response, err error) {
@\label{lst:deferAnonStart}@  defer func() {
@$\blacktriangleright$\label{lst:deferRace}@   resp, err = c.Foo(request, err)
  }()
  err = CheckRequest(request)
  ... // err check but no return
@\label{lst:deferGoStart}@  go func() {
@$\blacktriangleright$\label{lst:deferErrCapture}@   ProcessRequest(request, err != nil)
  }()
@\label{lst:ret2}@  return // the defer function runs after here
}  
\end{lstlisting}
\end{figure}
The anonymous function on line~\ref{lst:deferAnonStart} in Listing~\ref{lst:namedVarCaptureDefer} is prefixed with the \texttt{defer} keyword.
As a result, the anonymous function will execute after the return on line~\ref{lst:ret2}.
In this case, the developer has written the deferred function such that the two named return variables \texttt{resp} and \texttt{err} are correctly populated on any return path --- a defensive programming technique.

Unfortunately, the developer also uses a goroutine on line~\ref{lst:deferGoStart}, which captures the named return variable \texttt{err} on line~\ref{lst:deferErrCapture}.
When the execution returns on line~\ref{lst:ret2}, the write to the named return variable \texttt{err} on line~\ref{lst:deferRace} executes after the return. 
In the meantime, the read to the same \texttt{err} variable in the goroutine on line~\ref{lst:deferErrCapture} races with this write, causing a subtle data race. 
In the code where we found this situation, the enclosing function was large and complex.
The developer could not foresee the data race while developing the code because there is no visible concurrent access to \texttt{err} in the code after the launch of the goroutine.
Because of the complexity of the data race, the developer could not even understand the defect when our tool reported the issue. 

\subsection{Data Races due to Slices}

\begin{Observation}
Slices are highly confusing types that create subtle and hard to diagnose data races.
\end{Observation}

\begin{figure}[t!]
\begin{lstlisting}[language=Go,label={lst:interface},caption=Data race in slices even after using locks.]
func ProcessAll(uuids []string) {
@\label{lst:interfaceSliceDecl}@  var myResults []string
@\label{lst:interfaceMutexDecl}@  var mutex sync.Mutex
  safeAppend := func(res string) {
    mutex.Lock()
@$\blacktriangleright$\label{lst:interfaceAppend}@   myResults = append(myResults, res)
    mutex.Unlock()
  }

@\label{lst:interfaceLoop}@  for _, uuid := range uuids {
@\label{lst:interfaceGOLaunch}@    go func(id string, results []string) {
 @\label{lst:interfaceExpCall}@     res := Foo(id)
 @\label{lst:interfaceMyAppend}@     safeAppend(res)
@$\blacktriangleright$\label{lst:interfaceCall}@   }(uuid, myResults) // slice read without holding lock
  }
  ...
}
\end{lstlisting}
\end{figure}

Slices are dynamic arrays and reference types.
Internally, a slice contains a pointer to the underlying array, its current length, and the maximum capacity to which the underlying array can expand. 
We refer to these variables as \emph{meta} fields of a slice for ease of discussion.
A common operation on a slice is to grow it via the \texttt{append} operation, as shown on line~\ref{lst:interfaceAppend} in Listing~\ref{lst:interface}.
The slice is declared on line~\ref{lst:interfaceSliceDecl}.
When the size reaches the capacity, a new allocation (e.g., double the current size) is made, and the meta fields are updated.
When goroutines concurrently access a slice, it is natural to protect accesses to it by a mutex, as shown by the \texttt{Lock/Unlock} operations surrounding it.
In this code, the developer wraps the slice append operation into a standalone closure \texttt{safeAppend} and uses a mutex in the closure to ensure mutual exclusion; the closure correctly captures the free \texttt{mutex} variable.

A subtle bug happens because the developer accidentally passes the slice \texttt{myResults} as an argument as well to the goroutine, as shown on line~\ref{lst:interfaceCall}. This style of invocation causes the meta fields of the slice to be copied from the callsite to the callee.
Notice that this copying is \emph{not} lock protected.
While one iteration of the loop on line~\ref{lst:interfaceCall} is making a copy of the meta fields of \texttt{myResults}, a goroutine invoked on some previous iteration could be modifying the meta fields of the captured \texttt{myResults} on line~\ref{lst:interfaceAppend}. 
The \texttt{append} to this slice, which may modify the three meta fields of the slice, races with the read of the same meta variables.

This data race is not because of the goroutine capturing \texttt{myResults} by-reference, but because of a leftover passing of the slice. 
Given a slice is a reference type, one would not imagine its passing (copying) to a callee to cause a data race. 
However, a slice is not the same as a pointer type, hence the subtle data race. 
A better way to refactor this code is to a) pass a pointer to \texttt{myResults} and b) not capture \texttt{myResults} in the \texttt{safeAppend} closure, and instead arrange to dereference the incoming pointer to \texttt{myResults} while holding the lock.

\subsection{Data Races on Thread-Unsafe Map}
\begin{Observation}
Built-in maps in Go make them commonly used.
The array-style syntax of map accesses provides a false illusion of disjoint accesses of elements.
However, map implementation is thread-unsafe in Go causing frequent data races.
\end{Observation}

Hash table (\texttt{map}) is a built-in language feature in Go.
However, the built-in hash table in Go is not thread-safe.
Data race ensues if multiple goroutines simultaneously access the same hash table with at least one of them trying to modify the hash table (insert or delete an item).

Line~\ref{lst:mapSyntax} in Listing~\ref{lst:concurrentMap} shows the syntax for creating a hash table with a string key and an error value type.
The function \texttt{processOrders} receives a slice of \texttt{uuids}.
It iterates through each item in \texttt{uuids} (line~\ref{lst:mapIter}) and launches a goroutine per item to process each item asynchronously.
The \texttt{GetOrder} API invoked on each \texttt{uuid} may return a failure in which case the error string for the corresponding \texttt{uuid} is recorded in the \texttt{errMap} hash table. 
The developer intends to accumulate all errors (if any) in \texttt{errMap} and later return a combined error on line~\ref{lst:errCombine}.
A write-write data race happens on line~\ref{lst:mapRace} during concurrent writes to the hash table even though each \texttt{uuid} is intended to be a separate entry in the hash table.

In this case, the developer has full visibility into the concurrent code and the hash table (which is in the same enclosing function scope) and yet  makes a fundamental but common assumption (and a mistake) that the different entries in the hash table can be concurrently accessed. 
This assumption stems from developers viewing the  \texttt{errMap[uuid]} syntax used for map accesses and misinterpreting them to be accessing disjoint elements. However, a map (hash table), unlike an array or a slice, is a sparse data structure, and accessing one element might result in accessing another element; if during the same process another insertion/deletion happens, it will modify the sparse data structure and cause a data race.

\begin{figure}[t!]
\begin{lstlisting}[language=Go,label={lst:concurrentMap},caption=Concurrent access to hash map.]
func processOrders(uuids []string) error {
@\label{lst:mapSyntax}@  var errMap = make(map[string]error)
@\label{lst:mapIter}@  for _, uuid := range uuids {
    go func(uuid string) {
      orderHandle, err := GetOrder(uuid)
      if err != nil {
@$\blacktriangleright$\label{lst:mapRace}@      errMap[uuid] = err
       return
      }
      ...
    }(uuid)
@\label{lst:errCombine}@	return combineErrors(errMap)
}
\end{lstlisting}
\end{figure}

We found more complex concurrent map access data races (not shown) resulting from the same hash table being passed to deep call paths and developers losing track of the fact that these call paths mutate the hash table via asynchronous goroutines.

While the hash table leading to data races is not unique to Go, the following reasons make map-based data races common in Go.
\begin{description}[leftmargin=*]
    \item[Maps are more frequently used in Go:]
    Since map is a  built-in language construct, it is liberally used in Go.
    We have 83392 map-related constructs in our Java repository, whereas in Go, we have 273713 maps. 
This translates to  4389 map constructs per MLoC in Java, whereas the same for Go is 5950 per MLoC, which is $1.34\times$ higher. 
    \item[Array syntax and error tolerance:]
The map access syntax is just like array-access syntax (unlike Java's \texttt{get/put} APIs), making it easy to use but often confused for a random access data structure.
A non-existing map element can easily be queried with the \texttt{table[key]} syntax, which simply returns the default value without producing any error. This error tolerance can make one complacent when using Go maps.
\end{description}


\subsection{Mistakes due to Pass-by-Value vs. Pass-by-Reference in Go}
\begin{Observation}
Pass-by-value semantics are recommended in Go because it simplifies escape analysis and gives variables a better chance to be allocated on the stack, which reduces pressure on the garbage collector.
Developers often err on the side of pass-by-value (or methods over values), which can cause non-trivial data races.
\end{Observation}

\begin{figure}[t!]
\begin{minipage}{.9\linewidth}
\begin{lstlisting}[language=Go,label={lst:mutexValue},caption=Method invocation by value or pointer.]
var a int
// CriticalSection receives a copy of mutex.
@\label{lst:AStart}@func CriticalSection(m sync.Mutex) { 
@\label{lst:Lock}@  m.Lock()
@$\blacktriangleright$@ a++ 
  m.Unlock()
}
func main() {
@\label{lst:mutexCreate}@  mutex := sync.Mutex{}
  // passes a copy of m to A.
@\label{lst:mtexPass1}@  go CriticalSection(mutex)
@\label{lst:mtexPass2}@  go CriticalSection(mutex)
}
\end{lstlisting}
\end{minipage}
\begin{minipage}{.9\linewidth}
\begin{lstlisting}[language=Go,label={lst:syncMutexImpl},caption=sync.Mutex signature.]
type Mutex struct {
 // internal state
}
func (mtx *Mutex) Lock() {
 // lock implementation
}
func (mtx *Mutex) Unlock() {
 // unlock implementation
}
\end{lstlisting}
\end{minipage}
\end{figure}


Unlike Java, where all objects are reference types, in Go, an object can be a value type (struct) or a reference type (interface). There is no syntactic difference, leading to incorrect use of synchronization constructs such as \texttt{sync.Mutex} and \texttt{sync.RWMutex}, which are value types (structures) in Go.

In Listing~\ref{lst:mutexValue}, the function \texttt{main} creates a mutex structure on line ~\ref{lst:mutexCreate} and passes it by value to the goroutine invocation on lines~\ref{lst:mtexPass1} and~\ref{lst:mtexPass2}. 
The two concurrent executions of the function \texttt{CriticalSection}, now operate on two different mutex objects, which share no internal state.
As a result, the \texttt{Lock} and \texttt{Unlock} operations on these two different mutexes do not ensure mutually exclusive access to the global variable \texttt{a}.
A correct implementation should have passed the address of mutex (\texttt{\&mutex}), and the signature of \texttt{CriticalSection} should have received a pointer (\texttt{*sync.Mutex}).

Since the Go syntax is the same for invoking a method over pointers or values, less attention is given by the developer to question that \texttt{m.Lock} is working on a copy of \texttt{mutex} instead of a pointer.
Also, observe Listing~\ref{lst:syncMutexImpl} for how the \texttt{Lock} and \texttt{Unlock} are implemented.
Even though they (correctly) operate on a pointer to \texttt{Mutex}, the caller can still invoke these APIs on a \texttt{mutex} value, and the compiler transparently arranges to pass the address of the value.
Had this transparency not been there, the bug could have been detected as a compiler type-mismatch error.

A converse of this situation happens when developers accidentally implement a method where the receiver is a pointer to the structure instead of a value/copy of the structure.
In these situations, multiple goroutines invoking the method accidentally share the same internal state of the structure, whereas the developer intended otherwise.
Even in this case, the caller is unaware that the value type was transparently converted to a pointer type at the receiver.

\subsection{Mixing Shared Memory with Message Passing}
\begin{Observation}
Mixed use of message passing (channels) and shared memory makes code complex and susceptible to data races.
\end{Observation}
\begin{figure}[t!]
\begin{lstlisting}[language=Go,label={lst:channel},caption=Mixed use of shared memory and message passing.]
func (f *Future) Start() {
  go func() {
    resp, err := f.f() // invoke a registered function
@\label{lst:chanResprace}@    f.response = resp
@$\blacktriangleright$\label{lst:chanErrrace}@   f.err = err
@\label{lst:chanSend}@    f.ch <- 1 // may block forever!
  }()
}
func (f *Future) Wait(ctx context.Context) error {
@\label{lst:chanSelect}@ select {
@\label{lst:chanWait}@  case <-f.ch:
    return nil
@\label{lst:timeout}@   case <-ctx.Done():
@$\blacktriangleright$\label{lst:chanErrOnCancel}@  f.err = ErrCancelled
    return ErrCancelled
  }
}
\end{lstlisting}
\end{figure}

Listing~\ref{lst:channel} shows an example of a generic \texttt{Future}~\cite{JavafutureWWW} implementation by a developer using a channel for signaling and waiting.
The \texttt{Future} can be started by calling the \texttt{Start} method, and one can block for the future's completion by calling the \texttt{Wait} method on the \texttt{Future}.
The \texttt{Start} method creates a goroutine, which executes a function registered with the \texttt{Future} and records its return values (\texttt{response} and \texttt{err}).
The goroutine signals the completion of the \texttt{Future} to the \texttt{Wait} method by sending a message on the channel \texttt{ch} as shown on line~\ref{lst:chanSend}. 
Symmetrically, the \texttt{Wait} method blocks to fetch the message from the channel (line~\ref{lst:chanWait}).

\texttt{Contexts}~\cite{goContextWWW} in Go carry deadlines, cancelation signals, and other request-scoped values across API boundaries and between processes.
This is a common pattern in microservices where timelines are set for tasks.
Hence, \texttt{Wait} blocks on either the \texttt{context} being canceled (line~\ref{lst:timeout}) or the \texttt{Future} to complete (line~\ref{lst:chanWait}). 
Furthermore, the wait employs a \texttt{select} statement~\cite{golangChan} (line~\ref{lst:chanSelect}), which blocks until at least one of the \texttt{select} arms is ready.\footnote{if both are ready, one is chosen non-deterministically.} 

\sloppy
If the context times out, the corresponding case records the \texttt{err} field of \texttt{Future} as \texttt{ErrCancelled} on line~\ref{lst:chanErrOnCancel}. 
This write to \texttt{err} races with the write to the same variable in the future on line~\ref{lst:chanErrrace}.
Also, the goroutine will block forever on line~\ref{lst:chanSend} when there is no receiver on the other side of the channel.
Note that, if the context does \emph{not} timeout, which is the common case, the channel send operation establishes a happens-before edge preventing a data race.


\subsection{Incorrect Use of Flexible Group Synchronization}
\begin{Observation}
Go offers more leeway in its group synchronization construct  \texttt{sync.WaitGroup}. 
The number of participants is dynamic. 
Incorrect placement of \texttt{Add} and \texttt{Done} methods of a \texttt{sync.WaitGroup} lead to data races. 
\end{Observation}


\sloppy{}

The \texttt{sync.WaitGroup} structure is a group synchronization construct in Go.
Unlike C++ barrier~\cite{cppbarrier}, pthread barrier~\cite{ptbarrier}, and Java barrier~\cite{javabarrier} or latch~\cite{javalatch} constructs, the number of participants in a WaitGroup is not determined at the time of construction but is updated dynamically. Three operations are allowed on a \texttt{WaitGroup} object --- \texttt{Add(int)}, \texttt{Done()}, and \texttt{Wait()}.
\texttt{Add()} increments the \texttt{count} of participants, and the \texttt{Wait()} blocks until \texttt{Done()} is called \texttt{count} number of times (typically once by each participant). 
\texttt{WaitGroup} is extensively used in Go.
Table~\ref{tab:syncConstructs} shows that group synchronization is $1.9\times$ higher in Go than in Java.

In Listing~\ref{lst:incorrectAdd}, the developer intends to create as many goroutines as the number of elements in the slice \texttt{itemIds} and process the items concurrently.
Each goroutine records its success or failure status in \texttt{results} slice at different indices.
The developer expects the parent function block at line~\ref{lst:waitcorrect} until all goroutines finish.
The parent goroutine then accesses all elements of \texttt{results} to count the number of successful processings.

For this code to work correctly, when \texttt{Wait} is invoked on line~\ref{lst:waitcorrect}, the number of registered participants must be already equal to the length of \texttt{itemIds}.
This is possible only if \texttt{wg.Add(1)} is performed as many times as the length of \texttt{itemIds} prior to invoking \texttt{Wait}, which means \texttt{wg.Add(1)} should have been placed on line~\ref{lst:waitAdd}, prior to each goroutine invocation.
However, the developer incorrectly places \texttt{wg.Add(1)} inside the body of the goroutines on line~\ref{lst:waitAddWrong}, which is \emph{not} guaranteed to have been executed by the time the outer function \texttt{WaitGrpExample} invokes \texttt{Wait}. As a result, there can be fewer than the length of \texttt{itemIds} registered with the \texttt{WaitGroup} when the \texttt{Wait} is invoked.
For that reason, the \texttt{Wait} can unblock prematurely, and the \texttt{WaitGrpExample} function can start to read from the slice \texttt{results} (line~\ref{lst:waitResultRead}) while some goroutines are concurrently writing to it.

\begin{figure}[t!]
\begin{lstlisting}[language=Go,label={lst:incorrectAdd},caption=Incorrect WaitGroup.Add() function.]
func WaitGrpExample(itemIds []int) int {
 wg sync.WaitGroup
 results := make([]int, len(itemIds))
 for i := 0; i < len(itemIds); i++{
@\label{lst:waitAdd}@
   go(idx int) {
@\label{lst:waitAddWrong}@     wg.Add(1) // incorrect wg.Add placement
@$\blacktriangleright$@    results[idx] = ...
     wg.Done()
   }(i)
 }
@\label{lst:waitcorrect}@ wg.Wait() // waits for the participants added so far.
@$\blacktriangleright$\label{lst:waitResultRead}@... = results
}
\end{lstlisting}
\end{figure}
We also found data races arising from a premature placement of the \texttt{Done()} call on a \texttt{Waitgroup}. 

\subsection{Parallel Testing Idiom}
\begin{Observation}
Running tests in parallel for Go's table-driven test suite idiom can often cause data races, either in the product or test code.
\end{Observation}

Testing is a built-in feature in Go.
Any function with the prefix \texttt{Test} in a file with suffix \texttt{\_test.go} can be run as a test via the Go build system.
If the test code calls an API \texttt{testing.T.Parallel()}, it will run concurrently with other such tests.
We found a large class of data races happen due to such concurrent test executions.
The root causes of these data races were sometimes in the test code and sometimes in the product code.

Additionally, Go developers often write many subtests and execute them via the Go-provided \texttt{suite}~\cite{golangSuite} package within a single Test-prefixed function.
Go recommends a table-driven test suite idiom~\cite{CheneyTable} to write and run a test suite.
Our developers extensively write tens or hundreds of subtests in a test, which our systems run in parallel. 
This idiom becomes a source of problem for a test suite where the developer either assumed  serial test execution or lost track of using shared objects in a large complex test suite. 
Problems also arise when the product API(s) was written without thread safety (perhaps because it was not needed) but were invoked in parallel, violating the assumption.

\subsection{Incorrect or Missing Mutual Exclusion}
\begin{Observation}
Incorrect use of mutual exclusion primitives leads to data races.
\end{Observation}
This class of data races due to logical mistakes while writing concurrent programs is not unique to Go. It is also one of the most frequent reasons for data races in our code. We discuss this class of races for completeness. 

\subsubsection{Mutating shared data in a Reader-lock-protected critical section}

Go offers a reader-writer lock \texttt{RWMutex}, which allows concurrent readers to execute a critical section simultaneously. The \texttt{RLock/RUnlock} methods on a \texttt{RWMutex} hold the lock in a read-only mode. 
Sometimes developers accidentally put statements that may modify shared data in critical sections protected by \texttt{RWMutex}, while using \texttt{RLock/RUnlock} methods.
Such a mistake could have happened due to code evolving to address some issues or new requirements.
Listing~\ref{lst:RLock} shows one such example where after acquiring an \texttt{RLock}, the code may modify a shared variable on line~\ref{lst:RlockRace1}.
Concurrent readers may modify the same data leading to a data race.
Worse yet, on line~\ref{lst:RlockRace2}, the data race violates idempotency, where a network IO operation is performed more than once. 

\begin{figure}[t!]
\begin{lstlisting}[language=Go,label={lst:RLock},caption=Mutating a shared variable while holding a read-only lock.]
func (g *HealthGate) updateGate() {
  g.mutex.RLock()
  defer g.mutex.RUnlock()
  //... several read-only operations ...
  if ... {
@$\blacktriangleright$\label{lst:RlockRace1}@   g.ready = true // Concurrent writes.
@$\blacktriangleright$\label{lst:RlockRace2}@   g.gate.Accept() // More than one Accept().
  }
}
\end{lstlisting}
\end{figure}
\subsubsection{Other forms for incorrect mutual exclusion.}
There are numerous ways in which mutual exclusions can be incorrectly used.
The subtle ones we found were partial mutual exclusion, where the developer used locks in one place and forgot to use it in another while accessing the same shared variable(s).
In some cases, the developer used a lock but called unlock prematurely, leaving some shared variable access outside the critical section.
We observed an analogous situation where the developer used \texttt{sync.Atomic}~\cite{golangAtomic} partially --- used for writing to a shared variable but forgot to use it to read from the same variable.


\subsection{Summary of Findings}
\begin{table}[!t]
    \caption{Count of data races due to different Go language features and idioms.}
\footnotesize
    \centering
    \begin{tabular}{|c|l|c|}
\hline
Obs. \# & Description & Count \\ \hline
3 & Accidental capture-by-reference in a goroutine & 121 \\ \cline{2-3} 
& Capture-by-reference of err variable  & 50 \\ \cline{2-3}
& Capture-by-reference of loop range variable  & 48 \\ \cline{2-3}
& Capture of a named return  & 4 \\ \hline 
4 & Concurrent slice access & 391 \\ \hline 
5 &  Concurrent map access & 38 \\ \hline 
6 & Confusing pass-by-value vs. pass-by-reference & 38 \\ \hline     
7 & Mixing message passing with shared memory & 25 \\ \hline 
8 & Missing or incorrect use of group synchronization & 24 \\ \hline 
9 & Parallel test suite (table-driven testing) & 139 \\ \hline 
\end{tabular}
    \label{tab:categoryObser}
\end{table}

\begin{table}[!t]
    \caption{Count of data races due to language-agnostic reasons.}

\footnotesize
    \centering
    \begin{tabular}{|l|l|c|}
\hline
Note & Description & Count \\ \hline
Observation 10 & Missing or partial locking & 470 \\ \cline{2-3}  
& Mutating inside a reader-only lock & 2 \\ \hline 
Miscellaneous & Thread-safe APIs violating contract & 369 \\ \cline{2-3} 
 & Mutating a global variable & 24 \\ \cline{2-3} 
 &Missing or incorrect use of atomic ops & 40 \\ \cline{2-3}
 &Incorrect order of statements & 5 \\ \cline{2-3}
 &Complex multi-component interaction & 6 \\ \cline{2-3} 
 &Racy metrics / logging & 18 \\ \hline
Uncategorized &Fixed by removing concurrency & 26 \\ \cline{2-3} 
 &Fixed by disabling tests & 3 \\ \cline{2-3}
 &Fixed by a major refactor & 30 \\ \hline
\end{tabular}
    \label{tab:categoryMisc}
\end{table}

We studied each of the 1011 fixed data races and manually labeled their root cause(s).
Table~\ref{tab:categoryObser} shows the count of data races categorized under different language features and idioms discussed in this section.
Table~\ref{tab:categoryMisc} shows  the count of data races, which are language agnostic, and hence are not discussed in detail.
These labelings are not mutually exclusive; sometimes, multiple labels were assigned to the same bug because they fell into multiple categories.

Overall, we see that missing or incorrect locking (Table~\ref{tab:categoryMisc}) was the single largest reason for the data races.
Concurrent accesses to slices and maps were also common.
Capture-by-reference in goroutines can be seen to be the next common cause.
Mixing channels with shared memory, interchanging by-reference with by-value, and incorrect group communication are quite frequent causes of data races.
The other subtle causes grouped under the miscellaneous category are less common.
The last three lines in Table~\ref{tab:categoryMisc} pertain to those data races that were not root caused but instead addressed by refactoring the code: a) removing the concurrency, b)  disabling tests, or c) changing the code/logic in a significant way.

\begin{Remark}
    Future programming language designers should carefully weigh different language features and coding idioms with their potential to create common or arcane concurrency bugs.
\end{Remark}

\subsection{Threats to Validity}
The discussion in this paper is based on our experiences with data races in Uber's Go monorepo  and may have missed additional patterns of data races that may happen elsewhere. Also, as dynamic race detection does not detect all possible races due to code and interleaving coverage, our discussion may have missed a few patterns of races. 
Despite these threats to the universality of our results, the discussion on the patterns in this paper and the deployment experiences hold independently.

\section{Related Work}
Race detection is one of the most researched topics in the programming languages community with a vast body of literature. We discuss a few related works to help place our efforts in this context. 

\vspace{0.15cm}
\noindent \textbf{Dynamic race detection}: Dynamic analyses~\cite{fasttrack, djit, SchonbergDatarace, JohnOntheflyRace, FengCilkRace, YiziOpenMp, RaghavanRace, DinningDatarace} instrument code to capture memory accesses and synchronization events and reasons about them at runtime to  prove a conflict between a pair of accesses. Eraser~\cite{eraser}, which is based on the lock-sets algorithm, checks for empty lock-set intersection before accessing a shared variable to detect a race. Vector clocks~\cite{lamportClocks, djit, fasttrack} are commonly used in dynamic data race detection in unstructured parallel programs; ThreadSanitizer~\cite{SerebryanyThreadSanitizer}, which is based on these algorithms and used for race detection, can introduce 2$\times$-5$\times$ space and 2$\times$-20$\times$ runtime overheads.  
Race detection based on causally-precedes (CP) relation~\cite{cp-racedetection} improves precision at the cost of increasing the overheads further. 
Many subsequent enhancements have used strategies to reduce additional instrumentation~\cite{bond-oopsla15, hb-oopsla18}.  
While structured fork-join style parallel programs~\cite{openMPWWW, FrigoCilk, x10Allan} can significantly alleviate the overheads of runtime race detection~\cite{JohnOntheflyRace, FengCilkRace, YiziOpenMp, RaghavanRace}, this facility is not available in unstructured parallel programs such as Go.

Our study dives deeper into the deployment of a variant of dynamic race detection and the practical issues surrounding it. While any of the online approaches can be employed in our setup, the key problems of scalability and determinism of detecting races persist with all these approaches.


\vspace{0.15cm}
\noindent \textbf{Exploring thread interleavings}:
Executing multithreaded tests and observing assertion failures is yet another approach to help detect data races. As manifesting a race is dependent on the thread interleaving, {\sc RaceFuzzer}~\cite{racefuzzer} fuzzes the thread schedules by inserting random thread yields at different synchronization points to ensure different interleavings are explored. In contrast, {\sc Chess}~\cite{chess} systematically explores various thread interleavings by performing a tree traversal on the interleaving tree. 
Finally, TSVD~\cite{tsvd} uses lightweight monitoring of calling behavior associated with thread-unsafe methods to inject delays to expose races. 
While these strategies can also be used instead of dynamic analysis for our deployment, the problem of non-determinism with the detected races and the scale of the overall state space poses its own challenges for practical deployment.

\vspace{0.15cm}
\noindent \textbf{Automated test generation}:
A key requirement for dynamic analysis or testing-based approaches is the ability to analyze concurrent executions, which necessitates the availability of effective test suites.  Samak et al. leverage sequential execution traces to generate tests to expose data races~\cite{samak-pldi15}, other types of concurrency bugs~\cite{samak-oopsla14, samak-fse15}, and assertion violations ~\cite{samak-oopsla16}. In ~\cite{pldi12-pradel}, Pradel and Gross develop a technique for automatically generating executions with random inputs and thread interleavings to expose thread safety violations. Billes et al.,~\cite{pldi17-pradel} use a two-phase approach to identify potential conflicts by analyzing single client executions and synthesizing multi-client interactions. 
Our current deployment can benefit from all these approaches as they build the necessary test corpus for concurrent execution, which the dynamic race detector can analyze. 

\vspace{0.15cm}
\noindent \textbf{Surveys on concurrency bugs}: Tu et al.,~\cite{tu-asplos19} analyze 171 reported concurrency-related bugs in six open-source Go applications. They classify the bugs into blocking and non-blocking categories and provide the root causes in each case. Qin et al.,~\cite{qin-pldi2020} studied memory and thread-safety practices in real-world Rust programs, indicating the potential for data races in Rust programs where compile time checks are circumvented. More than a decade ago, Lu et al.,~\cite{lu-asplos08} studied four open-source C/C++ applications and characterized concurrency bug  patterns in them. 
In contrast to these  studies, our study 
a) focuses on data race bugs in Go,
b) explores significantly more (2100) services inside a real industrial setting, 
c) investigates more than 1000 data races, 
d) bubble-up the interplay between language features and data races, and 
e) discusses the software engineering aspects of deploying a dynamic race detection for industrial scale.

While Lu et al.,~\cite{lu-asplos08} found user-defined synchronization a common culprit in concurrency bugs, the same does not hold good for Go programs.
Go developers rarely, if at all, use their own synchronization but liberally use Go's Mutex locks and condition variables.

\vspace{0.15cm}
\noindent \textbf{Static analysis based race detection}: Static analysis for data race detection inspects source code to report conflicting accesses to shared variables. RacerD~\cite{racerd} performs abstract interpretation to report concurrency issues, including data races. {\sc Chord}~\cite{chord} uses various static analysis strategies for Java programs, including alias analysis, lock analysis, thread-escape analysis, and call-graph construction, to reduce the possible pairs of memory accesses involved in a data race.  Errorprone~\cite{errorprone} employs AST analysis to detect possible data races. MC checker~\cite{coverity} also uses compiler extensions to perform abstract interpretation and statistical analysis to report locations of ineffective locking. We believe the bug patterns in Go presented in this paper can inspire further research in static race detection for Go.  

\vspace{0.2cm}
\noindent \textbf{Detecting races in event-driven applications}: Petrov et al.~\cite{PetrovWebRacer} explore data races in JavaScript web applications 
by developing a definition of happen-before for asynchronous event-based web execution environments. Hsiao et al.~\cite{HsiaoMobileRace} explore data races in Android apps and 
observe that traditional dynamic data race detectors introduce false ordering between events handled sequentially by the same thread.   While Go has event-based asynchronous processing via channels, it is subsumed under the task-level parallelism of goroutines, where the traditional happens-before via vector clocks suffices. Further, unlike these efforts which explored novel race detection algorithms, the focus of this paper is on the deployment challenges and the interplay of language design with concurrency in Go.

\section{Conclusions}
In this paper, we discussed the characteristics of concurrency in Go programs and a deployment strategy for detecting races continuously using dynamic race detection. Based on observed (including fixed) data races, we elaborate on the Go language paradigms that make it easy to introduce races in Go programs. We hope that this work will inspire future work towards enabling dynamic race detection during continuous integration, developing program analyses for detecting Go races, and designing language features such that their interplay with concurrency does not easily introduce races. 

\bibliography{ref}


\end{document}